\documentclass[aps,pre,reprint,longbibliography,amsmath,amsfonts,amssymb,groupedaddress,floatfix]{revtex4-2}
\usepackage{hyperref}
\usepackage{graphicx}
\DeclareMathOperator{\Var}{Var}
\begin{document}

\title{Lattice model for percolation on a plane of partially aligned sticks with length dispersity}

\author{Avik P. Chatterjee}
\email{achatter@esf.edu}
\affiliation{Department of Chemistry, SUNY-ESF, One Forestry Drive, Syracuse, NY 13210, USA}
\affiliation{The Michael M. Szwarc Polymer Research Institute, Syracuse, NY 13210, USA}

\author{Yuri Yu. Tarasevich}
\email[Corresponding author: ]{ytarasevich@id.uff.br}
\affiliation{Instituto de F\'{\i}sica, Universidade Federal Fluminense, Niter\'oi, RJ, Brasil}
\date{\today}

\begin{abstract}
A lattice-based model for continuum percolation is applied to the case of randomly located, partially aligned sticks with unequal lengths in 2D which are allowed to cross each other. Results are obtained for the critical number of sticks per unit area at the percolation threshold in terms of the distributions over length and orientational angle and are compared with findings from computer simulations. Consistent with findings from computer simulations, our model shows that the percolation threshold is (i) elevated by increasing degrees of alignment for a fixed length distribution, and (ii) lowered by increasing degrees of length dispersity for a fixed orientational distribution. The impact of length dispersity is predicted to be governed entirely by the first and second moments of the stick length distribution, and the threshold is shown to be quite sensitive to particulars of the orientational distribution function. 
\end{abstract}

\maketitle

\section{Introduction\label{sec:intro}}

The formation of connected networks of microscopic particles (or other entities) that span macroscopic length scales comparable to the dimensions of the system they are embedded in is referred to as percolation and is related to substantial changes in material properties such as electrical conductivity~\cite{Balberg1983PRB,Borchert2015N,Carmona1987PRB,Du2005PRB,Majidian2017SR,Yook2012JKPS}. Examination of this subject using theoretical as well as computational methods has a deep history in the field~\cite{Balberg1983SSC,Yook2012JKPS,Chatterjee2014JCP,Tarasevich2018,Stauffer,Balberg2020}, and an understanding of conditions conducive to achieving percolation can assist in designing composites with desirable characteristics. Recent computer simulations have modeled effects due to varying the degree of length dispersity of the particles~\cite{Tarasevich2018,Bissett2023} thereby enabling a more realistic picture that accounts for the fact that the ideal situation of identically-sized particles is seldom encountered in experiments. The impact of varying degrees of alignment, which can reflect the conditions under which a material has been processed, has been examined as well in simulations for both two-dimensional (2D)~\cite{Yook2012JKPS,Klatt2017JSMTE,Tarasevich2018,Gotesdyner2022} and three-dimensional (3D) systems~\cite{Rahatekar2010}. The impact of these factors (length dispersity and degree of alignment) upon the percolation threshold is important for predicting the characteristics of thin film conductive structures based upon elongated particles such as carbon nanotubes or metal nanowires~\cite{Engel2008,Hicks2009PRE,Jagota2015,Sannicolo2016,Ackermann2016SR,Jagota2020,Dong2017,Redondo2022,Hu2023}. In this study we examine the application of a lattice-based theoretical approach~\cite{Chatterjee2012,Chatterjee2014} for modeling continuum percolation to the situation of partially aligned sticks with unequal lengths that are distributed randomly upon a planar 2D surface.

Our formalism exploits an analogy between lattice and continuum percolation that is based upon preserving the average number of contacts that lead to connectedness between individual entities. It is a method that has been employed with some success in capturing the impact of length dispersity upon the percolation threshold for rod-like particles in 3D~\cite{Nigro2013}, and which has been shown to accurately describe the dependence of the percolation threshold upon aspect ratio for isotropic, equally sized, penetrable rectangles in 2D~\cite{Chatterjee2014}. In this study we obtain a compact analytical expression for the critical areal density of sticks (rectangles of zero width) at the percolation threshold that accounts in a simple way for distributions over both the lengths of the sticks and their orientational angles. Results from this model are compared with those obtained from computer simulations for the appropriate orientational distributions that were employed in those investigations~\cite{Yook2012JKPS,Tarasevich2018}. Consistent with the simulations, we find that broadening the length distribution (for a fixed average stick length, $\langle L \rangle$) and increasing alignment have antagonistic effects upon the percolation threshold: the former lowers the threshold, while the latter elevates it. 

Section~\ref{sec:model}  presents our theoretical methodology and the different angular probability distribution functions (PDFs) that we use as illustrative examples and for comparison with simulations. Results from our comparison with computer simulations are presented with our conclusions in Section~\ref{sec:results}. 	Section~\ref{sec:concl} summarizes our main findings.

\section{Model for percolation by  penetrable rectangles with unequal lengths\label{sec:model}}
\subsection{Lattice-based model for continuum percolation\label{subsec:modlatt}}

We commence by considering a system of rectangles of uniform width (denoted $w$) and a distribution over lengths (denoted $L$) that are independent and identically distributed  upon a 2D planar surface. The PDF over lengths, denoted $f (L)$, satisfies 
\begin{equation}\label{eq:PDFnormaization}
  \int_0^\infty dL\, f(L) =1,
\end{equation}
and moments of the length distribution are denoted 
\begin{equation}\label{eq:moments}
\left\langle L^m \right\rangle = \int_0^\infty dL\, f (L ) L^m.
\end{equation}
The rectangles are assumed to be completely interpenetrable and are allowed to overlap, and we denote by $\rho$ the number of rectangles (of all lengths) per unit area. Each overlap between a pair of such (penetrable) rectangles creates a ``contact'' by virtue of which that pair of objects is defined to be ``connected''. The excluded area between a pair of rectangles of lengths $L_i$ and $L_j$ is denoted $A_{ij}$ and satisfies
\begin{equation}\label{eq:Aij}
A_{ij}= \left(L_i L_j+ w^2 \right)  |\sin\gamma |+w (L_i+ L_j )  (1+ |\cos\gamma |)  ,     
\end{equation}
for a given prescribed value of the angle $\gamma$ between the longitudinal axes of the pair of objects (\cite{Chatterjee2014} cf.~\cite{Balberg1984}). When averaged over the angular PDF for the selected pair of rectangles with lengths $L_i$ and $L_j$, Eqn.~\eqref{eq:Aij} leads to  
\begin{multline}\label{eq:meanAij}
\langle A_{ij}  \rangle = \left(L_i L_j+ w^2 \right)  \langle |\sin\gamma | \rangle \\ +w (L_i+ L_j )  (1+ \langle |\cos\gamma| \rangle ), 
\end{multline}
where we have assumed that the orientational distribution functions are identical for all values of the lengths so that the averages $\langle |\sin\gamma| \rangle$  and  $\langle |\cos\gamma| \rangle $ are independent of the lengths $L_i$ and $L_j$ of the pair of rectangles under consideration. (In the more general case in which the angular PDFs depend upon aspect ratio, the averages of $|\sin\gamma|$ and $|\cos\gamma|$ will depend upon the indices $i$ and $j$ and the terms appearing in~\eqref{eq:meanAij} must be denoted $\langle |\sin\gamma| \rangle _{ij}$ and $\langle |\cos\gamma| \rangle _{ij}$ instead.) The average number of contacts that a rectangle of length $L_i$ is expected to experience with all of the other rectangles in the system is denoted $n_{c,i}$ and can be estimated from 
\begin{equation}\label{eq:nci}
  n_{c,i}= \rho \int_0^\infty dL_j \,  f (L_j )  \langle A_{ij}  \rangle   .   
\end{equation}
Our heuristic mapping between the problems of (i) percolation by this system of rectangles and of (ii) percolation on a generalization of the Bethe lattice proceeds as follows~\cite{Chatterjee2012,Chatterjee2014}. We associate a vertex degree $z_i$ with each possible value of the length $L_i$ and the vertices of the lattice are assumed to be occupied with a uniform probability 
\begin{equation}\label{eq:xi}
  \xi = \rho w \int_0^\infty dL \,f (L) L = \rho w \langle L \rangle.
\end{equation}
The approach used in our study requires that the value of $\xi$ always be smaller than unity, and therefore applicability is restricted to the range $\rho < \left(w \langle L \rangle\right)^{-1}$. This requirement is satisfied for each of the cases we investigate and our calculated thresholds.

Each occupied site with vertex degree $z_i$ in the lattice model is intended to represent a rectangle of length $L_i$. The vertex degrees are related to the lengths by imposing the requirement that the average number of contacts be the same within the different models, that is 
\begin{equation}\label{eq:zi}
  z_i =  \frac{n_{c,i}}{\xi} =  \frac{1}{w \langle L \rangle } \int_0^\infty dL_j \, f (L_j )  \langle A_{ij}  \rangle.
\end{equation}
Equations~\eqref{eq:meanAij} and~\eqref{eq:zi} lead to
\begin{multline}\label{eq:zi1}
  z_i= \left(\frac{L_i}{w} + \frac{w}{\langle L \rangle  } \right) \langle |\sin\gamma| \rangle  \\+ \left(\frac{L_i}{\langle L \rangle}  +1\right)  (1+ \langle |\cos\gamma| \rangle ).    
\end{multline}
Within the assumption that closed loops can be neglected and that the lattice can be treated as being tree-like, our problem transforms to that of percolation on a Bethe lattice with a distribution over vertex degrees that reflects the distribution over lengths for the system of rectangles, and with a site occupation probability given by $\xi $. The critical site occupation probability at the percolation threshold for such a lattice, denoted $\xi _c$, is given~\cite{Cohen2000,Newman2001,Chatterjee2012} by 
\begin{equation}\label{eq:xic}
  \xi _c=  \frac{\langle z \rangle }{\langle z^2  \rangle - \langle z \rangle }, 
\end{equation}
where $\langle z \rangle$  and $\langle z^2  \rangle$  are the average values of the vertex degree and square of the vertex degree, respectively, averaged over all the vertices in the lattice. From~\eqref{eq:zi1} we find that
\begin{multline}\label{eq:meanz}
\langle z \rangle =  \frac{1}{w\langle L \rangle   } \\ \times\left[\left(\langle L \rangle ^2+ w^2 \right)  \langle |\sin\gamma| \rangle +2w\langle L \rangle   \left(1+ \langle |\cos\gamma| \rangle \right)\right]   ,    
\end{multline}
and 
\begin{multline}\label{eq:meanz2}
\langle z^2  \rangle 
= \left( \varepsilon^2 P + 2 + \varepsilon^{-2} \right)  \langle |\sin\gamma| \rangle ^2 
\\+ \left( P + 3\right)  (1+\langle |\cos\gamma| \rangle )^2 
\\ +2\left(\varepsilon P  + \varepsilon + 2 \varepsilon^{-1}  \right)  \langle |\sin\gamma| \rangle (1+\langle |\cos\gamma| \rangle),
\end{multline}
where
$$
P = \frac{\langle L^2  \rangle }{\langle L \rangle ^2}, 
$$
while  
$$
\varepsilon = \frac{\langle L \rangle}{w}
$$
is the aspect ratio.
Equations \eqref{eq:xic}, \eqref{eq:meanz}, and \eqref{eq:meanz2} yield 
\begin{equation}\label{eq:rhoc}
  \rho _c \langle L \rangle ^2=  \frac{\varepsilon  \langle z \rangle }{\langle z^2  \rangle - \langle z \rangle} = \varepsilon \xi_c,            
\end{equation}                                                                
where $\langle z \rangle$  and $\langle z^2  \rangle$   are given by  \eqref{eq:meanz} and \eqref{eq:meanz2}. It is straightforward but tedious to verify from  \eqref{eq:meanz} and \eqref{eq:meanz2} that $\langle z^2  \rangle >2 \langle z \rangle$  for any choice of the angular PDF and hence that the critical value of the site occupation probability $\xi _c$ is always smaller than unity, as it ought to be.  

The result in \eqref{eq:meanz}, \eqref{eq:meanz2},  and \eqref{eq:rhoc} simplifies considerably in the limit of vanishing widths (or very large aspect ratios), which we refer to as the ``stick limit''. If one takes the limit $w \to 0$ for fixed choices of the angular PDF and length distribution function $f(L)$, \eqref{eq:meanz}, \eqref{eq:meanz2},  and \eqref{eq:rhoc}  lead to
\begin{equation}\label{eq:rhocmeanL}
\rho _c^\text{stick} \langle L \rangle ^2=  \frac{P}{ \langle |\sin\gamma| \rangle}  =  
\frac{1}{(1+ \Sigma^2) \langle|\sin\gamma| \rangle },     
\end{equation}
where $\Sigma = \sigma _L/\langle L \rangle $, while $\sigma _L$ denotes the standard deviation in the lengths of the sticks. The factorization of the result in Ref.~\onlinecite{Chatterjee2012} into parts that depend exclusively upon (i) the length dispersity and (ii) the angular PDFs arises from our assumption that the angular PDFs are independent of the stick lengths. For an isotropic orientational distribution in two dimensions $\langle |\sin\gamma| \rangle =\langle |\cos\gamma| \rangle =  2/\pi $, and \eqref{eq:rhocmeanL} reduces to
\begin{equation}\label{eq:rhosiso}
  \rho _\text{c,iso}^\text{stick} \langle L \rangle ^2=  \frac{\pi }{2}P.  
\end{equation}
For the case of perfectly aligned rectangles for which $\gamma \equiv 0$, \eqref{eq:xic}, \eqref{eq:meanz},  \eqref{eq:meanz2}, and \eqref{eq:rhoc} yield
\begin{equation}\label{eq:rhocL2}
  \rho _\text{c,aligned} \langle L \rangle ^2= \frac{\varepsilon }{P + 2},
\end{equation}
which remains finite so long as the width $w$ is greater than zero. 

\subsection{Examples of Angular Distribution Function\label{subsec:APDF}}

Ascertaining the exact angular distribution function for elongated particles for a real experimental system is, in general, an arduous and difficult task~\cite{Trotsenko2015,Ackermann2016SR,Tomiyama2019,Hu2023}. Under most circumstances complete information regarding the angular distribution is not readily available. In view of this reality we have chosen a number of different angular PDFs to model possible orientational distributions. Within the assumption that the angular PDF is independent of the stick length we find that the dependence (as opposed to the absolute value) of the threshold upon length dispersity is independent of the particular PDF that is chosen and is governed by the scaled standard deviation of the stick length, namely $\Sigma$. Additionally, our model predicts that the threshold depends upon length dispersity exclusively through $\Sigma$ regardless of the distribution function describing the stick lengths.

As simple models for the state of alignment of the rectangles, we employ the four following illustrative angular PDFs $g (\theta)$, which we designate as angular PDFs A, B, C, and D.
 
\paragraph{Angular PDF A.}
This representation envisages the following step function structure for $g (\theta)$
\begin{equation}\label{eq:gA}
g (\theta)=  
\begin{cases}
  \dfrac{1}{2\alpha},  & \text{ for } |\theta|  \leqslant  \alpha ,  \\
  \\
  0 , & \text{ for } |\theta| >  \alpha , 
\end{cases}
\end{equation}
where $\alpha  \leqslant   \pi /2$ and the angular PDF is normalized such that 
$$
\int_{-\pi /2}^{\pi /2}d\theta \, g(\theta)=1.
$$ 
For this choice of angular PDF the orientational order parameter, denoted $S$, satisfies
\begin{equation}\label{eq:S}
  S= \langle \cos 2\theta \rangle =   \frac{\sin 2\alpha}{2\alpha},
\end{equation}
and the averages $\langle |\sin \gamma  |\rangle$  and  $\langle |\cos \gamma  |\rangle$  are given by \cite{Chatterjee2014}
\begin{equation}\label{eq:meansine}
  \langle |\sin \gamma  |\rangle = \frac{1}{\alpha } \left(1- \frac{\sin 2\alpha }{2\alpha} \right)  , 
\end{equation}
and 
\begin{equation}\label{eq:meancosine}
\langle |\cos \gamma  |\rangle = 
\begin{cases}
  \dfrac{1 - \cos 2\alpha }{2\alpha ^2 }, & \text{for } \alpha  \leqslant   \frac{\pi}{4},\\
  \\
  \dfrac{4\alpha - \pi + 1 +  \cos 2\alpha }{2\alpha ^2 }, & \text{for } \alpha  \geqslant   \frac{\pi}{4}. 
\end{cases}
\quad \end{equation}

\paragraph{Angular PDF B.}
This approximation was employed in prior investigations~\cite{Tarasevich2018,Gotesdyner2022} and envisages a normal (Gaussian) distribution over orientation angles as follows
\begin{equation}\label{eq:gtheta}
  g(\theta)=  \frac{1}{\sqrt{-\pi  \ln S }}  \exp\left(\frac{\theta^2}{\ln S} \right)  , 
\end{equation}
for which the variance in the angles $\theta$ is given by $\Var (\theta)= -0.5 \ln S$. For this choice of angular PDF, averages such as $\langle |\sin \gamma  |\rangle$  and  $\langle |\cos \gamma  |\rangle$  must be calculated by numerical integration. 

\paragraph{Angular PDF C.}
In this approximation the angular PDF is approximated by a pair of symmetrically distributed Dirac delta functions as follows 
\begin{equation}\label{eq:gthetaC}
  g (\theta)=  \frac{1}{2}  \left[\delta (\theta-\alpha )+ \delta (\theta+\alpha )\right],  
\end{equation}
where $\alpha$  is selected to reproduce desired values for the orientational order parameter $S$. For any prescribed value of $S$, angular PDF C has the smallest possible variance in the angle made by the longitudinal axes of the rectangles with the alignment direction, and we find that
\begin{equation}\label{eq:meansineC}
  \langle |\sin \gamma  |\rangle =  \frac{1}{2}  \sqrt{1- S^2 }    
\end{equation}
and
\begin{equation}\label{eq:meancosineC}
\langle |\cos \gamma  |\rangle = \frac{1}{2}  (1+ S).                                                                                                             \end{equation}

\paragraph{Angular PDF D.} 
In this approximation the angular PDF is represented by a pair of Dirac delta functions that are aligned parallel and perpendicular to the alignment direction
\begin{equation}\label{eq:gthetaD}
  g(\theta)= f_0  \delta (\theta)+ (1- f_0 )  \delta \left(\theta- \frac{\pi}{2}\right)  ,   
\end{equation}
where the weight factor $f_0$ is selected to reproduce desired values of the orientational order parameter $S$. For any prescribed value of $S$, angular PDF D has the largest possible variance in the angle made by the longitudinal axes of the rectangles with the alignment direction, and we find that
\begin{equation}\label{eq:meansineD}
  \langle |\sin \gamma  |\rangle =\frac{1}{2} \left(1- S^2 \right)   
\end{equation}
and
\begin{equation}\label{eq:meancosineD}
  \langle |\cos \gamma  |\rangle =  \frac{1}{2}  \left(1+ S^2 \right)  .  
\end{equation}
It should be noted that the choice $S = 0$ reduces to an isotropic distribution for PDFs A and B, but not for C and D. Our choices for angular PDFs are intended to be illustrative and are not by any means exhaustive, and other choices are possible and have been investigated~\cite{Klatt2017JSMTE,Tomiyama2019,Jagota2020}.

\section{Results\label{sec:results}}

In comparing our findings with published results of computer simulations, we focus attention on the case of rectangles with zero-width, namely, the stick limit \eqref{eq:rhocmeanL}. For the case of sticks that are equally sized and isotropically oriented, simulations have shown \cite{Li2013} that $\rho _c L^2=5.63724$  whereas our model yields $\rho _c L^2=  \pi /2$ for PDF A~\eqref{eq:gA}, while $\rho _c L^2=  1 /2$ for PDF C~\eqref{eq:meansineC} and PDF D~\eqref{eq:meansineD}. In order to compare various results in a way that examines the dependence (as opposed to the absolute value) of $\rho _c$ upon system variables, we multiply our result in \eqref{eq:rhocmeanL} by an appropriate factor, $\rho_0$,
\begin{equation}\label{eq:Resrho}
  \rho _c^\text{stick} \langle L\rangle ^2=  \frac{\rho_0}{(1+ \Sigma^2)\langle |\sin \gamma |\rangle}  .  
\end{equation}
Our choice for the factor $\rho_0$ depends on the particular choice of angular PDF. 
The choice of the prefactor in \eqref{eq:Resrho} enforces quantitative agreement between our model and the simulation-derived results for equally sized sticks for which $S=0$~\cite{Li2013}, and enables focusing upon relative variations in $\rho _c$ instead. 
For an isotropic system of equal-length sticks,  $\sigma _L^2=0$, $\langle L\rangle = L$, 
hence, 
\begin{equation*}\label{eq:Resrhomonoiso}
  \rho _c^\text{stick} L ^2=  \frac{\rho_0}{\langle |\sin \gamma |\rangle}.  
\end{equation*}
Equation \eqref{eq:meansine}  yields
$$
\langle |\sin \gamma |\rangle  
\propto
\frac{2}{3}\alpha, \text{ when } \alpha \to 0,
$$ 
hence, for the isotropic case,
\begin{equation}\label{eq:Resrhomonoiso1}
  \rho _c^\text{stick} L ^2=  \frac{3\rho_0}{2\alpha}.  
\end{equation}

\begin{figure}[!htb]
  \centering
  \includegraphics[width=\columnwidth]{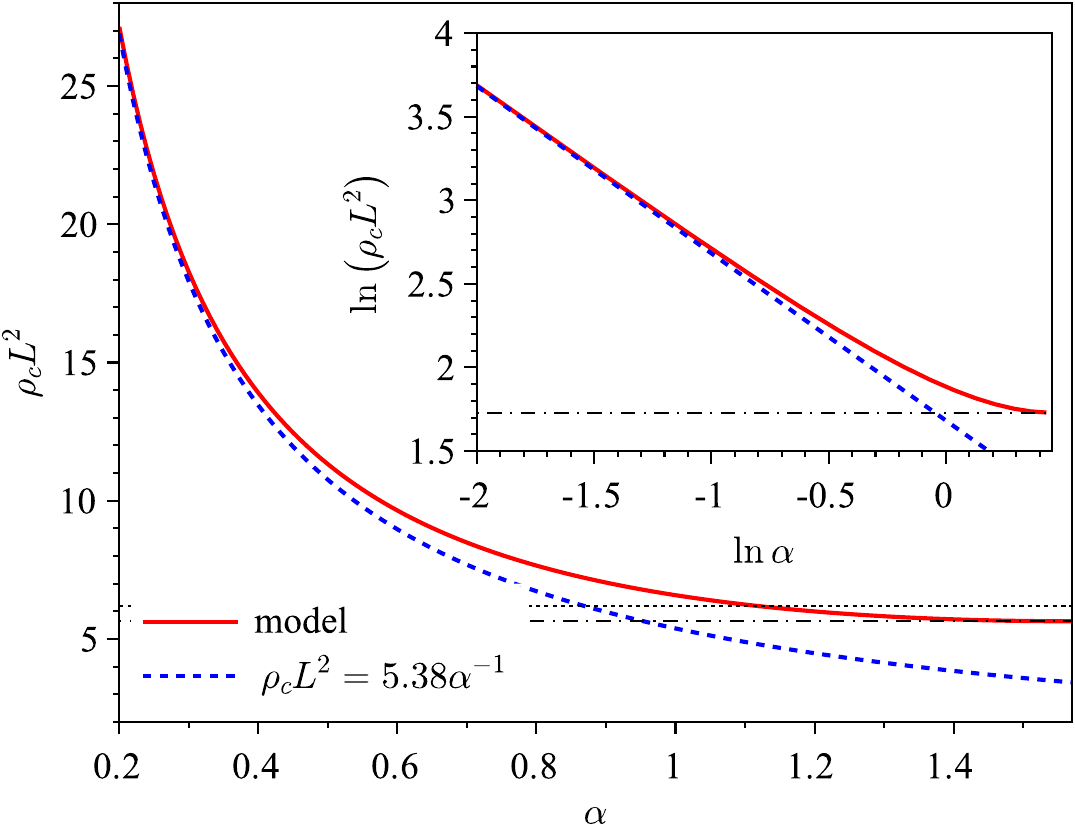}
  \caption{The solid line depicts the percolation threshold for equally sized, partially aligned sticks calculated from \eqref{eq:meansine} and \eqref{eq:Resrho} for angular PDF A \eqref{eq:gA} as a function of the cutoff angle~$\alpha$. The dashed line depicts the behaviors $\rho _c  L^2 = 5.38 \alpha^{-1}$~\eqref{eq:Resrhomonoiso1}. Dash-dot line corresponds to the percolation threshold for isotropic systems. The horizontal dotted line corresponds to the threshold for isotropic systems multiplied by the factor 1.1.\label{fig:Fig1} }
\end{figure}

The effect of partial alignment upon the percolation threshold for equally sized sticks for the stepfunction angular PDF (PDF A) is shown in Fig.~\ref{fig:Fig1}. The computer simulation study reported in \citet{Yook2012JKPS} found that for this choice of angular PDF (i) $\rho _c L^2$ depended only weakly upon the cutoff angle $\alpha$  for $\alpha  \geqslant   5\pi /18$, but that (ii) for $\alpha  <  5\pi /18$ the threshold increased with the degree of alignment, approximately following $\rho _c L^2   \propto  \alpha ^{-0.9}$. Our results are qualitatively similar to these findings with the distinction that \eqref{eq:meansine} and \eqref{eq:Resrho} predict that for high degrees of alignment ($\alpha \lessapprox 0.4$), $\rho _c  L^2 \propto \alpha^{-1}$  instead. The dashed line in Fig.~\ref{fig:Fig1} denotes the behavior $\rho _c  L^2 = 5.38 \alpha^{-1}$. For highly anisotropic systems ($\alpha \gtrapprox 1.1$), variation in the percolation threshold is less than 10\% (dotted line depicts the value $1.1\rho_c^\text{iso} L^2$). Our findings are qualitatively consistent with the observations reported in~\cite{Yook2012JKPS}.

\begin{figure}[!htb]
  \centering
  \includegraphics[width=\columnwidth]{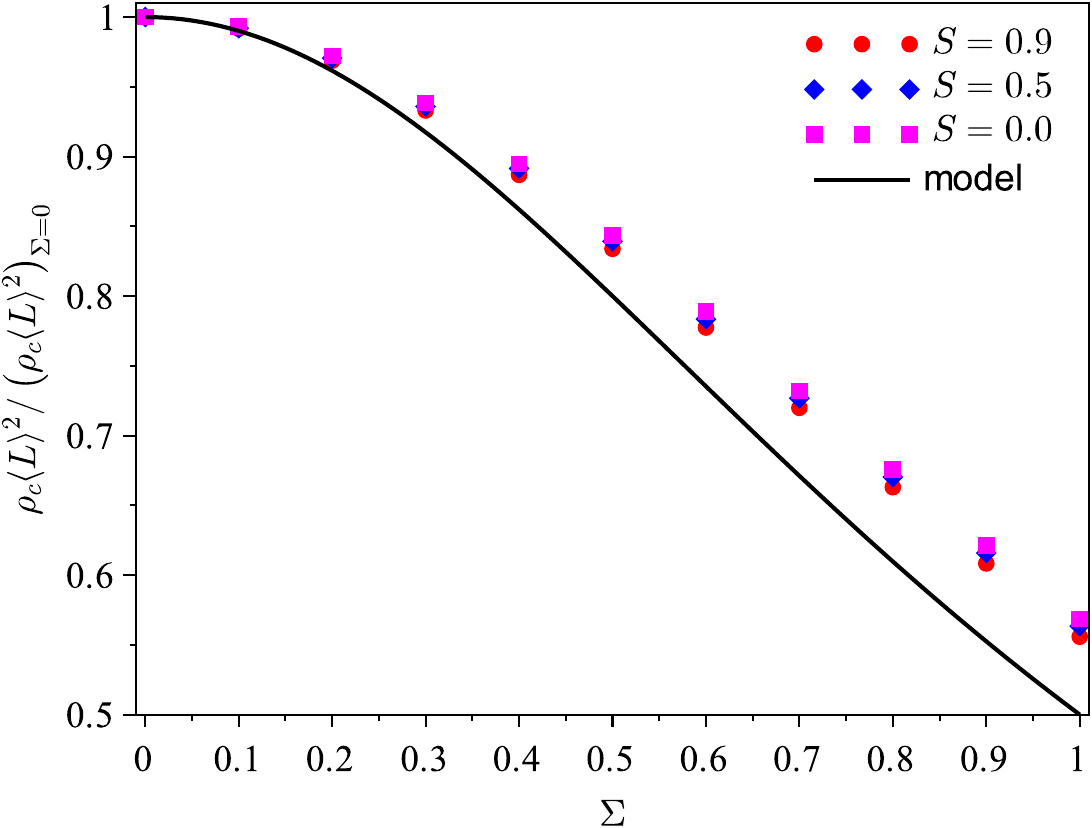}
  \caption{The percolation threshold $\rho _c \langle L\rangle ^2/ \left(\rho _c \langle L\rangle ^2\right)_{\Sigma=0}$ is shown as a function of the degree of length dispersity of the sticks ($\Sigma$) for fixed values of the orientational order parameter $S$. The markers represent the simulation results from Ref.~\onlinecite{Tarasevich2018} for values of $S= 0.9, 0.5, 0$. The line represents calculations from our model \eqref{eq:threshnorm}. \label{fig:Fig2}}
\end{figure}

The combined effects of both length dispersity and alignment upon the percolation threshold have been examined in the simulation study reported in~\cite{Tarasevich2018}, which investigated a log-normal distribution over stick lengths and a Gaussian distribution (angular PDF B in the present account) over stick angular orientations. Figure~\ref{fig:Fig2} depicts the dependence of the normalized percolation threshold
\begin{equation}\label{eq:threshnorm}
  \frac{\rho _c \langle L\rangle ^2}{\left(\rho _c \langle L\rangle ^2\right)_{\Sigma=0}} = \frac{1}{ 1 + \Sigma^2}
\end{equation} 
upon dispersity in the stick lengths for various prescribed, fixed value of the order parameter $S$. (The value of $\langle |\sin\gamma|\rangle$ for a given value of $S$ is different depending upon the PDF that is used.) The line represents calculations from our model~\eqref{eq:threshnorm}. Markers represent simulation results from~\cite{Tarasevich2018}) for $S = 0.9, 0.5, 0$. Increasing the dispersity in stick lengths (quantified by $\Sigma$) for any value of $S$ always lowers the threshold. The results from the simulation~\cite{Tarasevich2018} collapse onto a single curve for each value of $S$, thereby exhibiting the dependence upon $\Sigma$ predicted by our model.

\begin{figure}[!htb]
  \centering
  \includegraphics[width=\columnwidth]{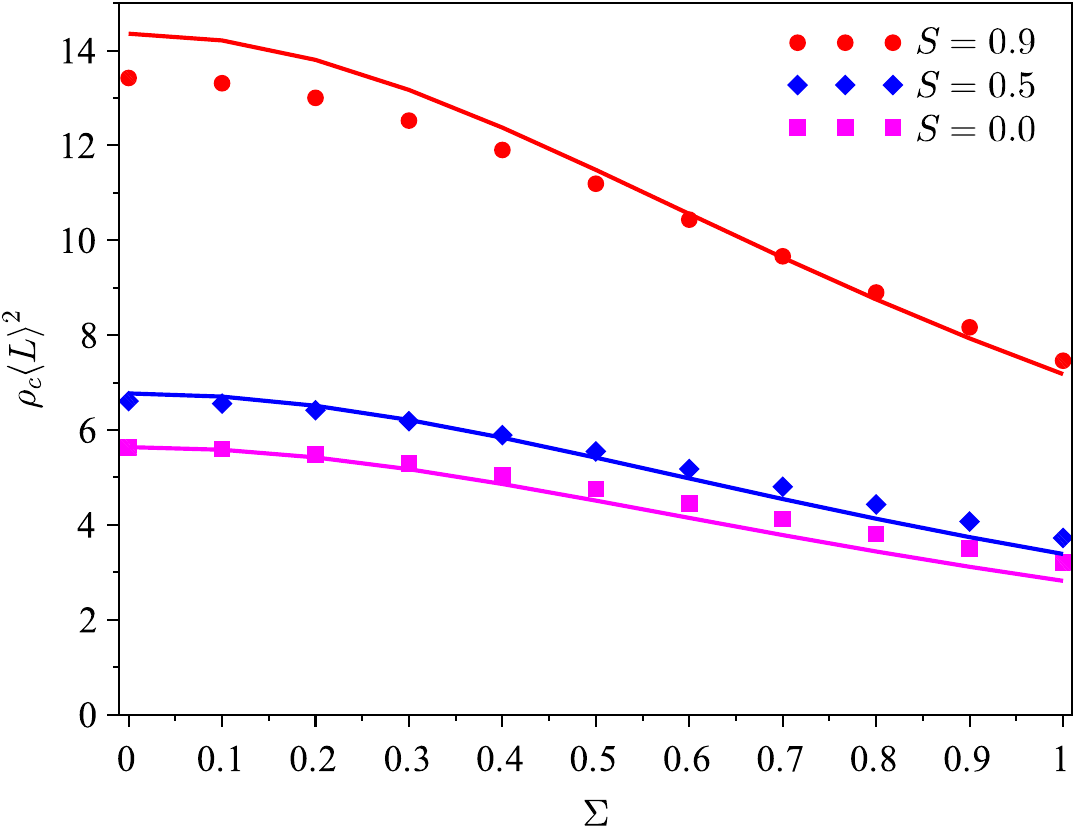}
  \caption{The percolation threshold $\rho _c \langle L\rangle ^2$ is shown as a function of the degree of lengths dispersity of the sticks ($\Sigma $) for different values of the orientational order parameter $S = 0.9, 0.5, 0$. The markers represent the simulation results from Ref.~\onlinecite{Tarasevich2018}. The  lines represent calculations from our model \eqref{eq:Resrho} using angular PDF B. \label{fig:Fig2Gauss}}
\end{figure}

Figure~\ref{fig:Fig2Gauss} reveals that results from the simulations of Ref.~\onlinecite{Tarasevich2018} are in close agreement with those obtained from our model using the same angular PDF (PDF B). The sensitivity of the predicted values of $\rho _c \langle L\rangle ^2$ to the choice of angular PDF for high degrees of stick alignment is demonstrated in Supplemental Material [URL will be inserted by publisher].  For the calculations depicted in Fig.~\ref{fig:Fig2Gauss}  that employ the Gaussian angular PDF (PDF B), the quantity $\langle |\sin \gamma |\rangle$  required in evaluating the right-hand-side of \eqref{eq:Resrho} is obtained by numerical integration. 
Increasing dispersity at a fixed value of $\langle L\rangle$  implies the presence of larger fractions of longer sticks that have a higher likelihood to cross and form contacts, thereby lowering the areal number density at the threshold. A comparable phenomenon has been reported for the case of (isotropic and random) rod-like particles in 3D, for which the weight-averaged aspect ratio was found to govern the percolation threshold~\cite{Nigro2013}. 

\begin{figure}[!htb]
  \centering
  \includegraphics[width=\columnwidth]{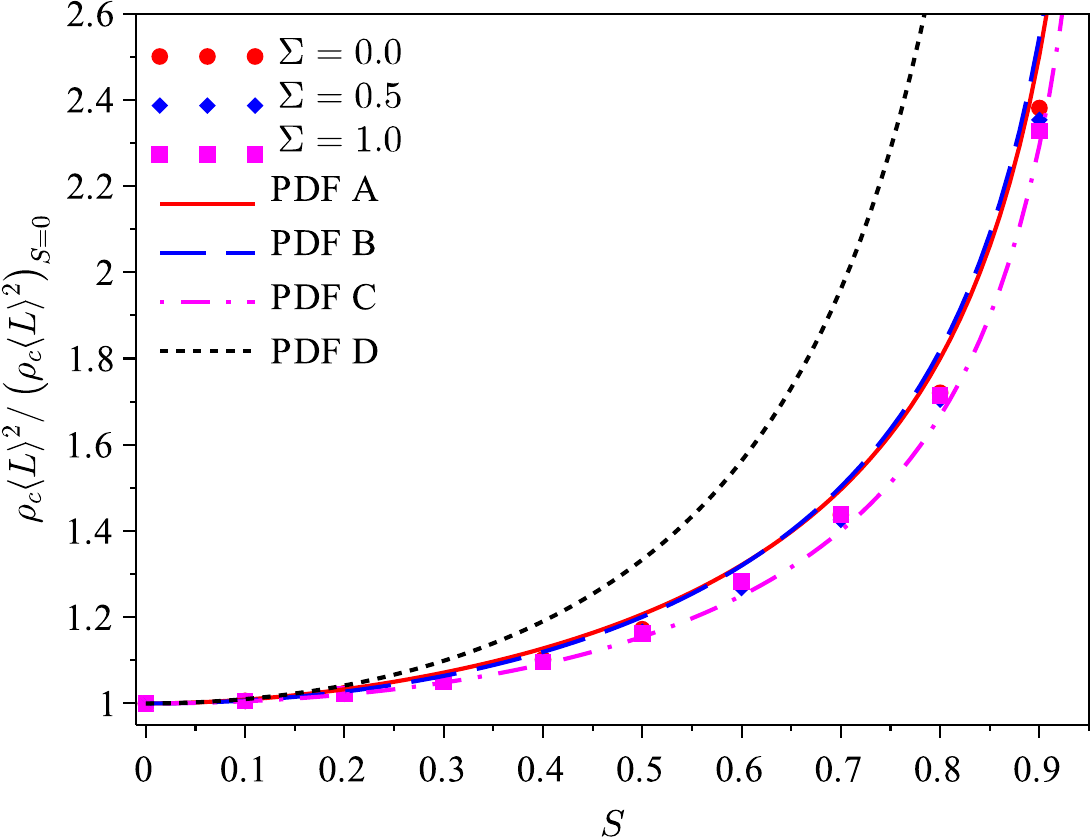}
  \caption{The percolation threshold $\rho _c \langle L\rangle ^2$ is shown as a function of the degree of alignment ($S$) of the sticks for fixed values of the dispersity in stick lengths ($\Sigma$). The diamonds, squares, and triangles represent the simulation results from \cite{Tarasevich2018} for $\Sigma = 1.0, 0.5, 0$, respectively. The solid, dashed, dash-dotted, and dotted lines represent calculations from our model \eqref{eq:Resrho} using angular PDFs A, B, C, and D, respectively.\label{fig:Fig3}}
\end{figure}

The dependence of 
\begin{equation}\label{eq:threshvsS}
\frac{\rho _c \langle L\rangle ^2}{\left( \rho _c \langle L\rangle ^2\right)_{S=0}} = \frac{\left.\langle|\sin\gamma|\rangle\right|_{S=0}}{\langle|\sin\gamma|\rangle}
\end{equation} 
upon the extent of alignment (quantified by $S$) for various fixed degrees of length dispersity  ($\Sigma$) is shown in Fig.~\ref{fig:Fig3}  for each choice of angular PDF. Consistent with the findings depicted in Fig.~\ref{fig:Fig1}, the threshold is seen to increase steeply with increasing degrees of alignment especially for large values of $S$. Results from our Gaussian PDF (PDF B) (the dashed line in Fig.~\ref{fig:Fig3}) are again similar to those reported in Ref.~\onlinecite{Tarasevich2018}. Results from the simulations~\cite{Tarasevich2018} and from model calculations for each PDF for different length dispersities collapse onto individual curves that depend only upon the value of $S$. (It should be noted that in our model $\rho _c \langle L\rangle ^2$ vanishes in the limits that either (i) $\langle L^2 \rangle /\langle L\rangle ^2 \to  \infty $, or (ii) $S  \to  0$.) Not surprisingly, the results from  model D are significantly different from the simulations because it is a very different PDF from that employed in the simulations. Results from Model D are intended solely to illustrate the possible variability in response to the specific choice of angular PDF \footnote{See Supplemental Material at [URL will be inserted by publisher] for additional figures for different PDFs.}.

\section{Discussion and conclusion\label{sec:concl}}

While the case of isotropic penetrable rectangles in 2D was examined in \cite{Chatterjee2014},  the present report examines the dependences upon length and angle dispersity.  \citet[Appendix B]{Balberg1984} suggested that length dispersity affects the threshold through the standard deviation of the lengths. Two possibilities (Eqns. B2 and B3) were mentioned as possibilities for the length disperse case. Comparison with simulation of a log-normal distributed system drew the authors of \cite{Balberg1984} to view their Eqn. B3 (which is identical to the result from our approach) as being more accurate. Our study derives this dependence from a model and confirms the proposed behavior. Additionally, our comparison with simulation results for partially aligned systems justifies and supports the dependence upon angular distribution suggested in \cite[Appendix B]{Balberg1984}  and gives a picture that can be applied to arbitrary angular distribution functions~\cite{Yook2012JKPS,Tarasevich2018PREb,Bissett2023}.

Our model qualitatively captures the trends seen in the simulations and the data collapse seen in Fig.~\ref{fig:Fig2}, however it predicts a stronger quantitative dependence upon $\Sigma$ than is seen in the simulations. This may arise due to variations in the importance of closed loops/cycles (which are entirely neglected in our approach) with the extent of length dispersity in a way that we cannot predict.

In concluding, we have extended a lattice-based model for continuum percolation to describe penetrable rectangles and zero-width sticks in 2D. Despite the fact that our approach neglects the impact of closed loops of connected objects, our results successfully describe major features of the dependence of the percolation threshold upon dispersity in the stick lengths and degree of alignment observed in computer simulation studies. In particular, the dependence upon the length distribution is confirmed to be governed by the first and second moments of the stick length through the ratio $\sigma _L/\langle L\rangle$   for any length distribution function, which is consistent with and generalizes existing findings for the log-normal distribution~\cite{Bissett2023,Tarasevich2018}. 
Our approach assigns a central role to the excluded area, and can be extended to other situations where estimates for this quantity are available.
For situations where only incomplete information is available about the length dispersity and angular PDFs, our study nevertheless identifies: (i) trends in the dependence of the threshold upon $S$ and $\Sigma$, and (ii) bounds (ascertained from the angular PDFs C and D) within which the threshold can vary for a prescribed value of $S$. The trends identified in our model are consistent with those reported from simulations~\cite{Yook2012JKPS,Tarasevich2018,Bissett2023}.

\begin{acknowledgments}
We acknowledge funding from the FAPERJ, Grants No.~E-26/202.666/2023 and No.~E-26/210.303/2023 (Y.Y.T.).
\end{acknowledgments}
\bibliography{dispersity}

\end{document}